\documentclass[aps, prl,twocolumn,superscriptaddress,showpacs,floatfix]{revtex4-1}
\usepackage{amsmath,amssymb,graphicx}
\usepackage{wasysym}
\usepackage[utf8]{inputenc}
\usepackage[T1]{fontenc}
\usepackage{xcolor}

\IfFileExists{newtxtext.sty}
   {\usepackage{newtxtext,newtxmath}}
   {\IfFileExists{stix.sty}
      {\usepackage{stix}}
      {\IfFileExists{mathptmx.sty}
      {\usepackage{mathptmx}}{} } }

\usepackage{textcomp}

\usepackage{bm}

\IfFileExists{siunitx.sty}{\usepackage{booktabs,siunitx}}{}

\pdfoutput=1
\usepackage{color}
\definecolor{LinkColor}{rgb}{0.256,0.439,0.588}
\usepackage{hyperref}

\renewcommand{\vec}[1]{\mathbf{#1}}

\newcommand{\Z}{\mathbb{Z}}

\usepackage{pifont}

\begin{document}

\title{Dynamical Signature of Symmetry Fractionalization in Frustrated Magnets}

\author{Guang-Yu Sun}
\affiliation{Beijing National Laboratory of Condensed Matter Physics and Institute of Physics, Chinese Academy of Sciences, Beijing 100190, China}
\affiliation{School of Physical Sciences, University of Chinese Academy of Sciences, Beijing 100190, China}
\author{Yan-Cheng Wang}
\affiliation{School of Physical Science and Technology, China University of Mining and Technology, Xuzhou 221116, China}
\author{Chen Fang}
\affiliation{Beijing National Laboratory of Condensed Matter Physics and Institute of Physics, Chinese Academy of Sciences, Beijing 100190, China}
\affiliation{CAS Center of Excellence in Topological Quantum Computation and School of Physical Sciences, University of Chinese Academy of Sciences, Beijing 100190, China}
\author{Yang Qi}
\affiliation{Center for Field Theory and Particle Physics, Department of Physics, Fudan University, Shanghai 200433, China}
\affiliation{State Key Laboratory of Surface Physics, Fudan University, Shanghai 200433, China}
\affiliation{Collaborative Innovation Center of Advanced Microstructures, Nanjing 210093, China}
\author{Meng Cheng}
\affiliation{Department of Physics, Yale University, New Haven, CT 06520-8120, U.S.A.}
\author{Zi Yang Meng}
\affiliation{Beijing National Laboratory of Condensed Matter Physics and Institute of Physics, Chinese Academy of Sciences, Beijing 100190, China}
\affiliation{CAS Center of Excellence in Topological Quantum Computation and School of Physical Sciences, University of Chinese Academy of Sciences, Beijing 100190, China}

\begin{abstract}
The nontrivialness of quantum spin liquid (QSL) typically manifests in the non-local observables that signify their existence, however, this fact actually casts a shadow on detecting QSL with experimentally accessible probes. Here, we provide a solution by unbiasedly demonstrating dynamical signature of anyonic excitations and symmetry fractionalization in QSL. Employing large-scale quantum Monte Carlo simulation and stochastic analytic continuation, we investigate the extended XXZ model on the kagome lattice, and find out that across the phase transitions from $\mathbb Z_2$  QSLs to different symmetry breaking phases, spin spectral functions can reveal the presence and condensation of emergent anyonic spinon and vison excitations, in particular the translational symmetry fractionalization of the latter, which can be served as the dynamical signature of the seemingly ephemeral QSLs in spectroscopic techniques such as inelastic neutron or resonance (inelastic) X-ray scatterings.
\end{abstract}

\date{\today}

\maketitle

{\it Introduction.-} Quantum spin liquids~\cite{BalentsSLReview,SLreview,LeonReview17,ZhouYi2017} are exotic phases of matter characterized by long-range many-body entanglement and fractionalized excitations~\cite{Wen2017}. One of the defining features of QSLs is that there is no local order parameter and the nontrivialness of the phase manifests in non-local observables. In the case of gapped QSL, global observables such as Wilson loop operators~\cite{SondhiWilsonloop2011, ShengPRL2005}, topological entanglement entropies~\cite{kitaev2006b, levin2006, Isakov2011} and modular transformations~\cite{zhang2012} have been exploited to characterize topological order theoretically, but none of them are directly accessible in experiments. Available to experiments are measurements of static~\cite{YCWang2017a} and dynamical spin correlation functions (besides thermodynamical quantities). In particular, dynamical spin structure factor (DSSF) measured by inelastic neutron scattering probes the spectral properties of elementary magnetic excitations~\cite{Han2012,WeiYuan2017}. It is therefore an important question to understand what kind of universal information about the underlying QSLs can be extracted from the momentum and energy resolved DSSFs.

\begin{figure}[htp!]
\centering	\includegraphics[width=0.9\columnwidth]{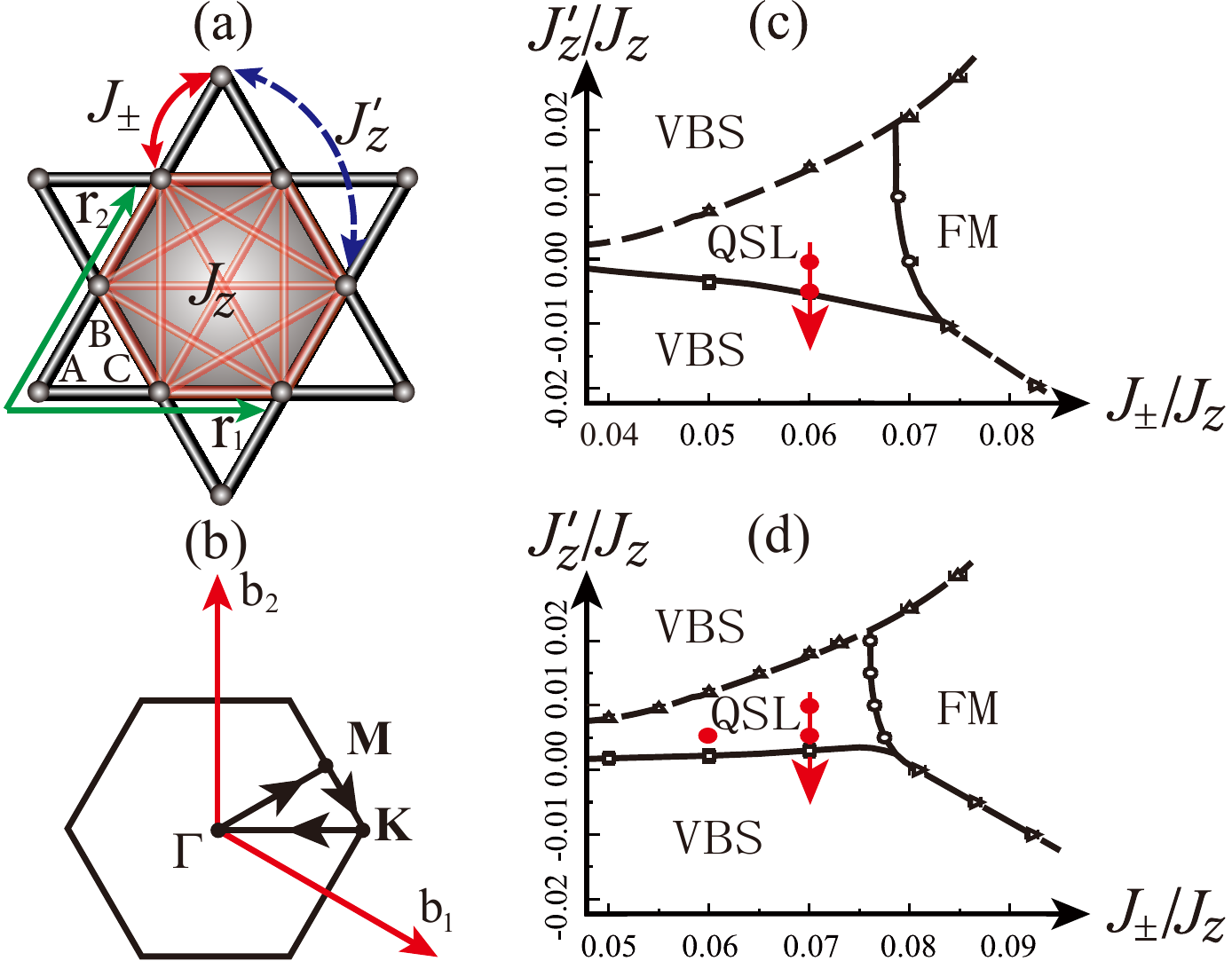}
	\caption{(a) The kagome lattice with lattice vectors $\mathbf{r}_1$ and $\mathbf{r}_2$, sublattices A, B and C, and all the interactions $J_{\pm}$, $J_z$ and $J'_z$ of the Hamiltonian in Eq.~(\ref{eq:hamiltonian}) depicted. (b) Brillouin zone (BZ) of the kagome lattice, with the reciprocal vectors $\mathbf{b}_1$ and $\mathbf{b}_2$, the high symmetry path goes through $\Gamma$, $M$ and $K$ points. (c) $J_{\pm}/J_z$ vs $J'_z/J_z$ phase diagram of Hamiltonian in Eq.~(\ref{eq:hamiltonian}) at magnetization $m_{z}=0$ ~\cite{YCWang2017a}. Along the $J_{\pm}/J_z$ axis, there is a transition from $\mathbb{Z}_2$ QSL to ferromagnetic ordered phase (FM); along the $J'_z/J_z$ axis, following the direction of the red arrow, there is a transition from $\mathbb{Z}_2$ QSL to a valence bond solid (VBS) phase which orders at $M$ point. (d) $J_{\pm}/J_z$ vs $J'_z/J_z$ phase diagram of Hamiltonian in Eq.~(\ref{eq:hamiltonian}) at magnetization $m_{z}=\frac{1}{6}$~\cite{YCWang2017b}. Along the $J_{\pm}/J_z$ axis, there is a transition from $\mathbb{Z}_2$ QSL to FM phase; along the $J'_z/J_z$ axis, following the direction of the red arrow, there is a transition from $\mathbb{Z}_2$ QSL to a VBS phase which orders at $\Gamma$ point.}
	\label{fig:LattPHD}
\end{figure}

For example, nowadays a continuum in DSSF is often taken as an indication of fractionalized excitations, but a simple continuum in the spin spectrum can also be caused by disorder~\cite{JSWen2018}.
Therefore, additional signatures in DSSF unique to a QSL are desired.
On the other hand, it is also desirable to read out more information besides the existence of fractionalized excitations from DSSF.
In particular, QSLs with the same type of anyon excitations can be further classified by how internal and lattice symmetries act on the anyons, known as the symmetry-enriched topological order~\cite{Barkeshli2014, Tarantino_SET, Wen2017}. It has been proposed that~\cite{wenpsg, EssinPRB2014, Mei2015} this additional information can also be detected from DSSF.
In this work, we for the first time compute the DSSF in a frustrated spin model in an unbiased manner and observe such unique dynamical signatures in DSSF -- the fractionalization of lattice symmetries -- in QSL with $\mathbb{Z}_2$ topological order.

\paragraph{Model and Method.-} We consider the extended Balents-Fisher-Girvin (BFG) model on a kagome lattice, where  $\mathbb{Z}_2$ QSLs are realized~\cite{BFG2002,Isakov2006,Isakov2011,Isakov2012,YCWang2017a,YCWang2017b,YCWang2018}. It has been extensively investigated as one of the very few models of frustrated magnets that can be simulated with unbiased quantum Monte Carlo (QMC) methods, and the defining features of QSL such as spinon and vison excitations~\cite{Isakov2006,Isakov2007}, topological entanglement entropy~\cite{Isakov2011} and fractionalized quantun critical point~\cite{Isakov2012} have been revealed.

The Hamiltonian of the model is given by
\begin{equation}
	\begin{split}
		H =&-J_{\pm}\sum_{\langle i,j \rangle} (S_i^{+}S_j^{-}+\text{h.c.}) + \frac{J_z}{2}\sum_{\hexagon}\Big(\sum_{i\in\hexagon} S_i^z\Big)^{2}\\
 &+J'_z\sum_{\langle i,j \rangle^{'}} S_i^{z}S_j^{z}-h\sum_{i}S_i^{z},
	\end{split}
\label{eq:hamiltonian}
\end{equation}
with the physical meaning of each term illustrated in Fig.~\ref{fig:LattPHD} (a). The original BFG model consists of the nearest-neighbour spin flip $J_{\pm}>0$ term, and $J_z>0$ plaquette interaction term for each hexagon. The newly added $J'_z$ term is a next-nearest-neighbour interaction that frustrates the ordering of spins on the same sublattice~\cite{YCWang2017b}. The Zeeman field $h$ tunes the total magnetization. We set $J_z=1$ as the energy unit. The Hamiltonian preserves all symmetries of the kagome lattice, as well as a $\mathrm{U}(1)_{S_z}\rtimes \mathbb{Z}_2=\mathrm{O}(2)$ spin symmetry at $h=0$ and the $\mathrm U(1)_{S_z}$ symmetry at $h\neq0$. Hence, throughout this paper, the spin quantum number refers to $S_z$.

\begin{figure*}[htp!]
	\centering	\includegraphics[width=0.9\textwidth]{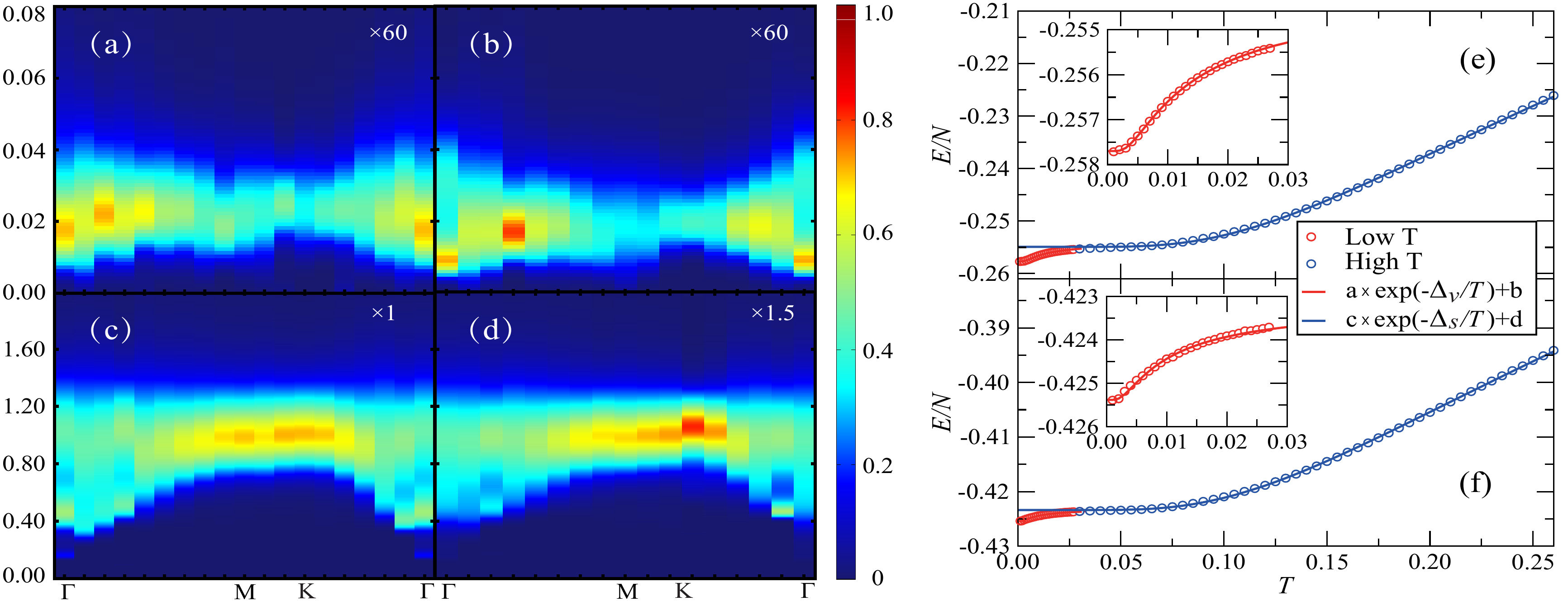}
	\caption{$S^{zz}(\mathbf{q},\omega)$ along the high symmetry path at $m_z=0$ with $J_{\pm}=0.06$, $J'_z=0$ (a) and $m_z=\frac{1}{6}$ with $J_{\pm}=0.06$, $J'_z=0.005$ (b). The system size is $L=16$. The spectra are all gapped with continua. The spectral bottom is at $\omega \sim 0.01$, this is the energy scale of a vison-pair as discussed in the text and consistent with the vison-pair gap ($\Delta_{v}$) fitted in (e) and (f). $S^{\pm}(\mathbf{q},\omega)$ along the high symmetry path at $m_z=0$ with $J_{\pm}=0.06$, $J'_z=0$ with $L=18$ (c)  and $m_z=\frac{1}{6}$ with $J_{\pm}=0.06$, $J'_z=0.005$ with $L=16$ (d). The spectra are all gapped with continua. The spectral bottom is at $\omega \sim 0.2$, consistent with the energy scale of a spinon-pair gap ($\Delta_s$) obtained in (e) and (f). (e) [(f)] is the temperature dependence of the energy for $L=12$ system for parameters in (a) and (c) [(b) and (d)], $\exp(-\Delta/T)$ fits are performed to extract anyonic gaps.}
	\label{fig:spectraSzSzSxSx}
\end{figure*}

To study the model in Eq.~(\ref{eq:hamiltonian}), we employ large-scale stochastic series expansion~\cite{Syljuaasen2002} QMC simulations. Since the model is highly anisotropic and frustrated, i.e., $J_{\pm} \ll J_z$ and $J'_{z}$, to avoid sampling problem of many local minima, we perform QMC simulation with a 5-spin plaquette update (10 legs in a vortex)~\cite{YCWang2017a,YCWang2017b}. Moreover, to reduce the rejection rate of the proposed spin configurations, we make use of algorithm that satisfies the balance condition without imposing detail balance~\cite{YCWang2017a,YCWang2017b,SuwaPRL2010}. The ground state phase diagrams of the model in Eq.~(\ref{eq:hamiltonian}) at $m_{z}=0 (\frac{1}{6})$ plateaus are determined in Refs.~\cite{YCWang2017a,YCWang2017b,YCWang2018}, reproduced in Fig.~\ref{fig:LattPHD} (c) and (d). In both cases, stable $\Z_2$ QSL emerges and the transitions from $\mathbb{Z}_2$ QSL to the FM phase driven by $J_{\pm}/J_z$ are continuous and of $(2+1)$ XY$^{*}$ universality~\cite{Isakov2012,YCWang2017b}.

\paragraph{Spinon and vison continua.-} Here we introduce the concept of spinon and vison excitations in the BFG QSL and develop an understanding of how they are detected in the DSSFs. We are primarily interested in the Ising limit $J_\pm\ll J_z$. The ground state manifold thus respects the constraint $S_{\hexagon}^z=\sum_{i\in\hexagon}S_i^z=6m_z$ for each hexagon. Violations of these constraints correspond to deconfined excitations, the spinon, whose energy gap is of the order of $J_z$. Since a spin flip $S^+_i$, which carries charge-$1$ under U$(1)_{S_z}$, creates two identical hexagon excitations each with $S_{\hexagon}^z=6m_z+1$, each of them must carry U$(1)_{S_z}$ charge-$1/2$ and thus is called a spinon. The other kind of excitation, visons, is more subtle and can be viewed as sources of $\pi$ flux for spinons. Namely, when a spinon is transported around a vison its wavefunction changes sign. Since visons do not carry any U$(1)_{S_z}$ charge, it is natural that $S^z$ operators can create pairs of visons. This is supported by an explicit construction of vison states at a soluble deformation of the BFG model in Ref. [\onlinecite{BFG2002}]. Because visons are created in the low-energy manifold with $S_{\hexagon}^z=6m_z$, they have a much lower energy gap of the order of $J_\pm^2/J_z$.

Therefore, to observe the spectral information of spinon and vison excitations of $\mathbb Z_2$ QSLs, we make use of the following dynamical spin structure factors:
\begin{eqnarray}
S^{\pm}_{\alpha\beta}(\mathbf{q},\tau)&=&\langle S^{+}_{-\mathbf{q},\alpha}(\tau)S^{-}_{\mathbf{q},\beta}(0) \rangle,\\
S^{zz}_{\alpha\beta}(\mathbf{q},\tau)&=&\langle S^{z}_{-\mathbf{q},\alpha}(\tau)S^{z}_{\mathbf{q},\beta}(0)\rangle.
\label{eq:corr_def}
\end{eqnarray}
Here, the imaginary time $\tau\in[0,\beta]$, and to make sure that the system is close to the QSL ground state, we choose $\beta = 2L/J_{\pm}$ to be below the energy scale associated with the anyonic excitation gap (if $J_{\pm}=0.1J_z$ then $T\le J^{2}_{\pm}/J_z$ when $L>5$). $L$ is the linear system size and the total number of sites $N=3\times L^2$. $\alpha,\beta=1,2,3$ label the three sublattices of the kagome lattice. $\langle\cdots\rangle$ stands for the QMC ensemble average. We have defined $S^{\pm}_{\mathbf{q},\alpha}=\sqrt{3/N}\sum_{i\in\alpha}e^{-i\mathbf{q}\cdot\mathbf{r}_i}S^{\pm}_{i}$, where the summation is over the sublattice $\alpha$ and $\mathbf{r}_i$ is the spatial position of the site $i$. $S^{zz}_{\mathbf{q},\alpha}$ is defined in the same vein.

One can obtain the real frequency DSSFs via the stochastic analytic continuation(SAC)~\cite{Sandvik1998a,Beach2004,Syljuasen2008,Sandvik2015,Qin2017,Shao2017a,Shao2017b,Huang2017} of the imaginary-time data. In SAC, candidate real-frequency spectra are proposed and fitted to the imaginary time data. Each candidate is then weighted by their goodness-of-fit $\chi^2$ as an effective energy such that a Metropolis sampling can be defined over the proposed spectra. The final spectrum is the ensemble average of all candidates. Detailed account of SAC and
its recent applications can be found in Refs.~\cite{Sandvik2015,Qin2017,Shao2017a,Shao2017b,Huang2017,YRShu2017}.

Fig.~\ref{fig:spectraSzSzSxSx} shows the obtained $S^{\pm}(\mathbf{q},\omega)=\frac{1}{3}\sum_{\alpha}S^{\pm}_{\alpha\alpha}(\mathbf{q},\omega)$ and $S^{zz}(\mathbf{q},\omega)=\frac{1}{3}\sum_{\alpha}S^{zz}_{\alpha\alpha}(\mathbf{q},\omega)$ along the high symmetry path. Fig.~\ref{fig:spectraSzSzSxSx} (a) and (c) are for $m_z=0$ with $J_{\pm}=0.06$, $J'_z=0$ and (b) and (d) are for $m_z=\frac{1}{6}$ with $J_{\pm}=0.06$, $J'_z=0.005$. The parameters are chosen such that the system is well inside the QSL phases, according to the phase diagrams in Fig.~\ref{fig:LattPHD} (c) and (d). It is clear that in both $S^x$ and $S^z$ channels,  the spectra are gapped with continua above the gap. As explained earlier, we expect that the magnon spectra observed in $S^{\pm}$ are comprised of spinon pairs to match the quantum number $S_z=1$, while the excitations in $S^{zz}$ are comprised of vison pairs to match with $S_z=0$. Since the spinons and visons are gapped, their pair spectra in $S^{\pm}(\mathbf{q},\omega)$ and $S^{zz}(\mathbf{q},\omega)$ are gapped as well. 

We can confirm this interpretation by examing the energy scales of the spectral gap. We expect that the spinon excitations have a pair gap $\Delta_s$ of the order of $J_z=1$, and for visons it is $\Delta_v\sim J^2_{\pm}/J_z$. From Fig. \ref{fig:LattPHD} we see that the value of $J_{\pm}/J_z$ for the onset of the QSL phases is $J_{\pm}/J_z \sim 0.1$, so we can estimate $\Delta_v\sim 0.01$.
 In Fig.~\ref{fig:spectraSzSzSxSx} (a)-(d), we find that the minimum of spectrum is located at the $\Gamma$ point for both spinons and visons, and we can read off $\Delta_s \approx 0.2$, and $\Delta_v \approx 0.01$, consistent with expectation.
 

Such consistency in energy scales can also be observed directly from the temperature dependence of the total energy. Fig.~\ref{fig:spectraSzSzSxSx} (e) and (f) show the $E(T)$ for QSLs at both $m_z=0$ with $J_{\pm}=0.06$, $J'_z=0$ and $m_z=\frac{1}{6}$ with $J_{\pm}=0.06$, $J'_z=0.005$ for $L=12$ system. It is clear that there are two different exponential decays in the curves, as shown by our $\exp(-\Delta/T)$ fitting, the higher energy gap is at $\Delta_s \sim 0.42$, which we identify as the spinon-pair gap. It is larger but of the same magnitude with that observed in the $S^{\pm}(\mathbf{q},\omega)$ in Fig.~\ref{fig:spectraSzSzSxSx} (c) and (d), and it is expected since the thermodynamic measurements contains the contribution from the full spectra hence give rise to a larger gap. And the second, much lower $\exp(-\Delta_v/T)$ happens at $\Delta_v \sim 0.01$ (see insets for clarity), which is apparently the vison-pair gap, consistent with the gap in $S^{zz}(\mathbf{q},\omega)$ in (a) and (b) in Fig.~\ref{fig:spectraSzSzSxSx}. Therefore, we can conclude that the vison-pair excitations are observed in the $S^{zz}(\mathbf{q},\omega)$ spectra and the spinon-pair excitations are observed in the $S^{\pm}(\mathbf{q},\omega)$ spectra. Similar observation for the vison excitation in pure BFG model at $m_z=0$ has also been shown in Ref.~\cite{Isakov2007}.

\paragraph{Symmetry fractionalization.} Once established the relation between DSSF and anyonic excitation gaps, we now set out to explore the more salient yet fundamental difference between the $\mathbb{Z}_2$ QSLs
at $m_z=0$ and $\frac{1}{6}$, i.e., their different form of symmetry fractionalizations~\cite{QiYang2015,QiCSF,QiSFV2015X}.
In the case of $\mathbb{Z}_2$ QSLs, spinons and visons can carry fractional crystal momentum associated with the lattice, which means that
\begin{equation}
	T_1^{(a)}T_2^{(a)}=-T_2^{(a)}T_1^{(a)},
	\label{eq:TT=-TT}
\end{equation}
here $a$ refers to an anyon (i.e. spinon or vison) and $T_{1,2}^{(a)}$ denotes the local action of translation on it. Intuitively, with such fractionalization, $a$ moves on a lattice with $\pi$-flux per unit cell.

The fractional crystal momentum carried by visons is determined by the magnetization per unit cell~\cite{ChengLSM}. In $\mathbb Z_2$ QSLs realized in the extended BFG models, the spinon carries a half-integer spin, a fractioanlized quantum number~\cite{BFG2002}.
At $m_z=0$, the spin per unit cell is $\frac32$, indicating that there must be an odd number of spinon therein.
As a result, moving a vison around a unit cell results in a $\pi$ Berry phase due to its mutual braiding with the spinons. In other words, the visons must carry a fractional crystal momentum and consequently vison translation operators anticommute at $m_z=0$. On the other hand, at $m_z=\frac{1}{6}$, the spin per unit cell is $1$, indicating that there must be an even number of spinon therein. Hence, the visons do not carry a fractional crystal momentum and the vison translation commutes~\footnote{Detials of the vison translational symmetry fractionalization, the consequential enhanced periodicity of density of states in DSSF, and the finite size effects in the obtained $S^{zz}(\mathbf{q},\omega)$ from the QMC+SAC simulations, are presented in the supplemental materials (SM).}. There will be two major differences of such fractional crystal momentum in the vison-pair continua at $m_z=0$ and $\frac{1}{6}$.

First, consider the limit where the spetral edge in DSSF is dominated by scattering states of a pair of visons.
If the vison carries a fractional crystal momentum, the density of the scattering states $\mathcal{N}(\vec{q}, \omega)$ should exhibit an enhanced periodicity~\cite{EssinPRB2014, Mei2015}:
\begin{equation}
	\mathcal{N}(\vec{q}, \omega) = \mathcal{N}(\vec{q}+\vec{K}, \omega),
	\label{eqn:Nperiodicity}
\end{equation}
where $\vec{K}$ is half of the reciprocal vector, i.e. $2\vec{K}=\vec{G}$. In particular, such enhanced periodicity should manifest in the spectral edge.

\begin{figure}[htp!]
\centering
\includegraphics[width=\columnwidth]{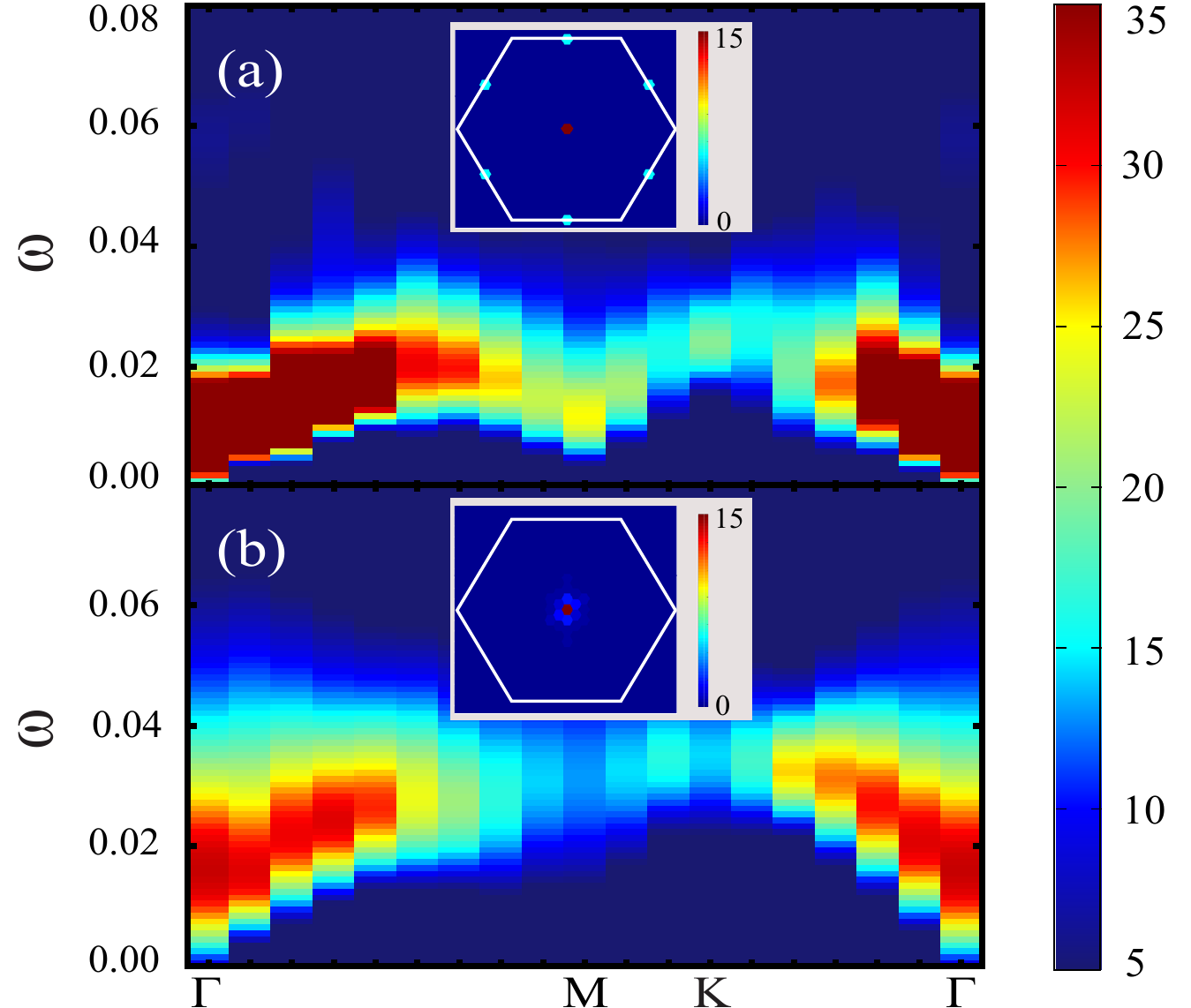}
\caption{The comparison of the vison-pair spectra close to QSL-VBS transition for $m_z=0$ with $J_{\pm}=0.06$, $J'_z=-0.005$ (a) and $m_z=\frac{1}{6}$ with $J_{\pm}=0.07$, $J'_z=0.005$ (b). The system size is $L=18$. In both cases, $S^{zz}(\mathbf{q},\omega)$ are becoming gapless due to the vison condensation. However, in (a), the enhanced periodicity manifests in that both $\Gamma$ and $M$ points become gapless, signifying of translational symmetry fractionalization, whereas in (b), there is no enhanced periodicity and hence no translational symmetry fractionalization. Insets show the static structure factor of the two VBS phases for $m_z=0$ and $\frac{1}{6}$, respectively, in (a), the Bragg peaks are at both $\Gamma$ and $M$, whereas in (b), only at $\Gamma$.}
\label{fig:periodvison}
\end{figure}

Second, the translational symmetry fractionalization is also reflected on the gap-closing momenta near the phase transition driven by vison condensation. Phase transitions between the QSL and nearby symmetry-breaking phases can be understood as driven by anyon condensations: the transition to the FM phase is driven by spinon condensation, and that to the VBS phases, both at $m_z=0$ and $\frac{1}{6}$, are driven by the vison condensation.
When the vison carries a fractional crystal momentum, its condensation will lead to the spontaneous breaking of the translational symmetry in the VBS phase.
Moreover, the fractional crystal momentum has the following constraints on the ordering wave vector $\vec q$ of the VBS phase: i) in the symmetry-breaking phase the static structure factor must be peaked at $\vec{q}+\vec{K}$ as well (the peaks do not have the same heights), ii) if we approach the critical point from the QSL side,  we expect to see gap closing in the DSSF at $\vec{q}$ and $\vec{q}+\vec{K}$. Both statements easily follow from the discussion of enhanced periodicity in the density of states of spin excitations if we treat the condensation transition at a ``mean-field'' level, but they hold more generally even when the interactions between visons can not be ignored.

To detect the translational symmetry fractionalization in the QSLs realized in our model, we monitor how the spectra evolve as the transition to the VBS is approached from the QSL side, indicated by the two paths in Fig.~\ref{fig:LattPHD} (c) and (d). These results are shown in Fig.~\ref{fig:periodvison}.
At $m_z=0$, the envelope of the two-vison continuum shows a enhanced periodicity between $\vec q$ and $\vec q+\vec K$, where the gap closes at both $\Gamma$ and $M$.
Correspondingly, the static structure factor in the VBS phase is found to have two peaks at $\Gamma$ and $M$ (see the inset of Fig.~\ref{fig:periodvison} (a)).
These features are consistent with the expectation that the vison carries a fractional crystal momentum.
On the other hand, at $m_z=\frac{1}{6}$, the envelope of the two-vison continuum is asymmetric between $\Gamma$ and $M$, the gap closes only at $\Gamma$, and the static structure factor in the VBS phase has only one peak at $\Gamma$ (see the inset of Fig.~\ref{fig:periodvison} (b)), consistent with the expectation that the vison does not carry a fractional crystal momentum at $m_z=\frac{1}{6}$.

The sharp contrast in Fig.~\ref{fig:periodvison} between $m_z=0$ and $\frac{1}{6}$, both in static and dynamic structure factors, clearly demonstrate the presence/absence of the translation symmetry fractionalizaiton in $\mathbb{Z}_2$ QSLs at the two magnetizations. This is, to our knowledge, for the first time being observed in non-perturbative manner. These results point out the possibility, that, inelastic neutron scattering or resonance (inelastic) X-ray scattering experiments can be further employed to identify gapped QSL on kagome magnets, for example in ZnCu$_3$(OH)$_6$Cl$_2$
(Herbertsmithite)~\cite{Han2012,Fu2015} and
Cu$_3$Zn(OH)$_6$FBr (Zn-doped Barlowite)~\cite{ZLFeng2017a,WenXG17,WeiYuan2017,ZLFeng2017b,Pasco2018,Ranjith2018,ZLFeng2018}. In both cases existing experimental data are pointing towards gapped QSL ground states with possibly
$\mathbb{Z}_2$ topological order, especially the latter, in which a gapped spinon continuum has been consistently revealed from both NMR~\cite{ZLFeng2017a} and inelastic neutron scattering experiments~\cite{WeiYuan2017}. If, when driving the material through a transition from QSL to ordered state, the doubled period could be observed in DSSF, such dynamical signature of symmetry fractionalization will be the decisive information to confirm the $\mathbb{Z}_2$ QSL in materials.

{\it Discussion.-} Employing large-scale QMC+SAC simulations, we study the DSSF in extended BFG models at different magnetizations.
We associate the DSSF of $S^zS^z$ and $S^+S^-$ operators with the two-particle continua of the vison and spinons.
The two-vison continuum reveals the difference in translation-symmetry fractionalization between QSLs at different $m_z$: at $m_z=0$, the continuum has an enhanced periodicity relating $\Gamma$ and $M$ points, meaning that vison carries the symmetry fractionalization. In contrast, at $m_z=\frac{1}{6}$, the continuum has no such enhanced periodicity, meaning that vison carries a trivial symmetry fractionalization instead.
Furthermore, at $m_z=0$, this enhanced periodicity also implies that the condensation of visons must occur simultaneously at $\Gamma$ and $M$, and thus breaks translation symmetries. Correspondingly, the condensation of visons at $m_z=\frac{1}{6}$ leads to a translational symmetric VBS state~\cite{YCWang2017b}.
Therefore, the dispersion of the two-vison continuum observed in the DSSF, together with the nature of the vison-condensation transition, reveals the symmetry fractionalization pattern of  the anyonic excitations in QSL.

Our findings show that the DSSF can not only detect the existence of fractionalized anyonic excitations in a QSL, but also distinguish different symmetry-fractionalization patterns carried by the anyons.
Since DSSF can be measured in different experimental probes, our findings hence not only have theoretical importance in understanding the properties of topological state of matter, but also provide a valuable experimental guide to look for dynamical signature of symmetry-enriched topological order in QSL materials.

{\it Acknowledgement.-} We would like to thank Stefan Wessel for communicating results of a related investigation~\cite{Becker2018} prior to publication and Anders Sandvik for valuable discussions. GYS, YCW and ZYM acknowledge fundings from the Ministry of Science and Technology of China through National Key Research and Development Program under Grant No. 2016YFA0300502, from the key research program of the Chinese Academy of Sciences under Grant No. XDPB0803 and from the National Science Foundation of China under Grant Nos. 11421092, 11574359 and 11674370 as well as the National Thousand-Young Talents Program of China. YQ is supported by the Ministry of Science and Technology of China under Grant No. 2015CB921700. MC is supported by startup funds from the Yale University. We thank the Center for Quantum Simulation Sciences in the Institute of Physics, Chinese Academy of Sciences, the Tianhe-1A platform at the National Supercomputer Center in Tianjin for their technical support and generous allocation of CPU time.

\bibliography{anyoncondensation}
\end{document}


\title{Supplymentary Materials for "Dynamical Signature of Symmetry Fractionalization in Frustrated Magnets"}

\maketitle

\section{Translational symmetry fractionalization}

\subsection{Filling constraint on symmetry fractionalization of visons}
\label{sec:vison-frac}
We review the derivation of the filling constraint in a $\Z_2$ QSL outlined in the main text, following Ref.[42]. We will focus on the U(1)$_{S_z}$ symmetry. It will turn out convenient to represent the spin-$1/2$'s as hard-core bosons: $S_z=b^\dag b-\frac{1}{2}, S^+=b^\dag, S^-=b$, and U(1)$_{S_z}$ symmetry becomes the boson number conservation. The relation between the average magnetization $m_z$ (per site) and the filling fraction $\nu$ of bosons is $\nu=3m_z+\frac{1}{2}$.

Suppose the ground state is a $\Z_2$ QSL, with the bosonic spinon $e$ carrying $\frac{1}{2}$ U(1) charge, vison $m$ neutral under U(1) and the fermionic spinon $\psi=e\times m$. Imagine adiabatically inserting a $2\pi$ flux in the $\Z_2$ QSL state. Here a flux means a point-like extrinsic defect, obtained by modifying the Hamiltonian. By definition, taking an excitation with charge $q$ counterclockwise around a flux $\phi$ results in the Aharonov-Bohm phase $e^{iq\phi}$. Flux is only defined mod $2\pi$.

First we show that a $2\pi$ flux must be identified with the $m$ anyon, i.e the vison. A $2\pi$ flux is gauge-equivalent to zero flux, so it should be identified with one of the anyon excitations, say $x$.  The statistical phase of braiding $a$ around $x$, denoted by $M_{a,x}$, is just $e^{2\pi iq_a}$, i.e. the Aharonov-Bohm phase. Given the assumed charge values of anyons, we obtain
\begin{equation}
	M_{e,x}=-1, M_{m,x}=1.
	\label{}
\end{equation}
This uniquely fixes $x=m$.

Now we consider the translation symmetry fractionalization of $m$. As we discussed in the main text, $T_1T_2T_1^{-1}T_2^{-1}$ acting on $m$ is equivalent to move $m$ around a unit cell. Since $m$ is identified with the $2\pi$ U(1) flux, moving the flux around the unit cell measures the average charge in the unit cell, which is $\nu$. We thus obtain
\begin{equation}
	T_1^{(m)}T_2^{(m)} = e^{2\pi i\nu}T_2^{(m)}T_1^{(m)}=-e^{6\pi im_z}T_2^{(m)}T_1^{(m)}.
	\label{}
\end{equation}
Therefore, if $m_z=0$ ($\frac{1}{6}$), $T_1$ and $T_2$ anticommute (commute) on the vison $m$.

\subsection{Enhanced periodicity for density of states}
We quickly review the derivation of Eq.(5) in the main text.
Intuitively, such enhanced periodicity can be viewed as a direct consequency of the $\pi$-flux per unit cell in the hopping amplitudes of the vison: a doubled unit cell is required to accomodate such $\pi$-flux, and therefore the dispersion has an enhanced periodicity in the momentum space.

To be more precise, we consider the scattering states for a pair of $a$, say $a_1$ and $a_2$, labeled by $\ket{\vec{k}_1, \vec{k}_2}$, where $\vec{k}_1+\vec{k}_2=\vec{q}$.  It is an eigenstate of the translation operator:
\begin{equation}
T_i \ket{\vec{k}_1, \vec{k}_2} = T_i^{(a_1)}T_i^{(a_2)} \ket{\vec{k}_1,\vec{k}_2}=e^{i\vec{q}\cdot\vec{e}_i}\ket{\vec{k}_1,\vec{k}_2}.
\label{}
\end{equation}
We now act on the state by the local translation operator on one of the anyons, say $T_1^{(a_1)}$. In the limit where the two anyons do not interact, this action creates a state degenerate in energy, but because of the anticommutation relation,
\begin{equation}
	T_1^{(a_1)}T_2^{(a_1)} = -T_2^{(a_1)} T_1^{(a_1)},
	\label{}
\end{equation}
we have
\begin{equation}
	\begin{split}
	T_2T_1^{(a_1)}\ket{\vec{k}_1,\vec{k}_1}&=T_2^{(a_1)}T_2^{(a_2)}T_1^{(a_1)}\ket{\vec{k}_1,\vec{k}_2}\\
	&=-T_1^{(a_1)}T_2^{(a_1)}T_2^{(a_2)}\ket{\vec{k}_1,\vec{k}_2}\\
	&=-T_1^{(a_1)}T_2\ket{\vec{k}_1,\vec{k}_2}\\
	&=-e^{i\vec{q}\cdot\vec{e}_2}T_1^{(a_1)}\ket{\vec{k}_1,\vec{k}_2}
	\end{split}
	\label{}
\end{equation}
which implies that the $T_2$ eigenvalue of $T_1^{(a_1)}\ket{\vec{k}_1,\vec{k}_2}$ is $-e^{i\vec{q}\cdot\vec{e}_2}=e^{i(\vec{q}+\frac{\vec{G}_2}{2})\cdot \vec{e}_2}$. It is obvious that $T_1$ eigenvalue of $T_1^{(a_1)}\ket{\vec{k}_1,\vec{k}_2}$ is still $e^{i\vec{q}\cdot\vec{e}_1}$.
Hence, for every state $\ket{\vec k_1,\vec k_2}$ with momentum $\vec q$ and energy $\omega$, there is another state, $T_1^{(a_1)}\ket{\vec k_1,\vec k_2}$, with the same energy $\omega$ but a different momentum $\vec q+\frac{\vec G_2}{2}$.
Thus we have shown that
		\begin{equation}
			\mathcal{N}(\vec{q},\omega)=\mathcal{N}\left(\vec{q}+\frac{\vec{G}_2}{2},\omega\right).
			\label{}
		\end{equation}
		Clearly the same argument holds for any half reciprocal vectors:
\begin{equation}
	\begin{split}
			\mathcal{N}(\vec{q},\omega)&=\mathcal{N}\left(\vec{q}+\frac{\vec{G}_1}{2},\omega\right),\\
			\mathcal{N}(\vec{q},\omega)&=\mathcal{N}\left(\vec{q}+\frac{\vec{G}_1+\vec{G}_2}{2},\omega\right)
	\end{split}
			\label{}
		\end{equation}
Therefore the density of states has an enhanced periodicity in the Brillouin zone.

Notice that both Ref.[20] and Ref.[21] focused on the enhanced periodicity in spinon-pair spectra, which only apply to certain types of ground states. In this paper we instead consider fractionalization of visons.
As shown in Sec. I. A., for $m_z=0$ visons necessarily have fractional momentum and thus vison-pair spectra should have enhanced periodicity.

\section{Finite size analysis of vison gap}

\begin{figure}[htp!]
\centering
\includegraphics[width=\columnwidth]{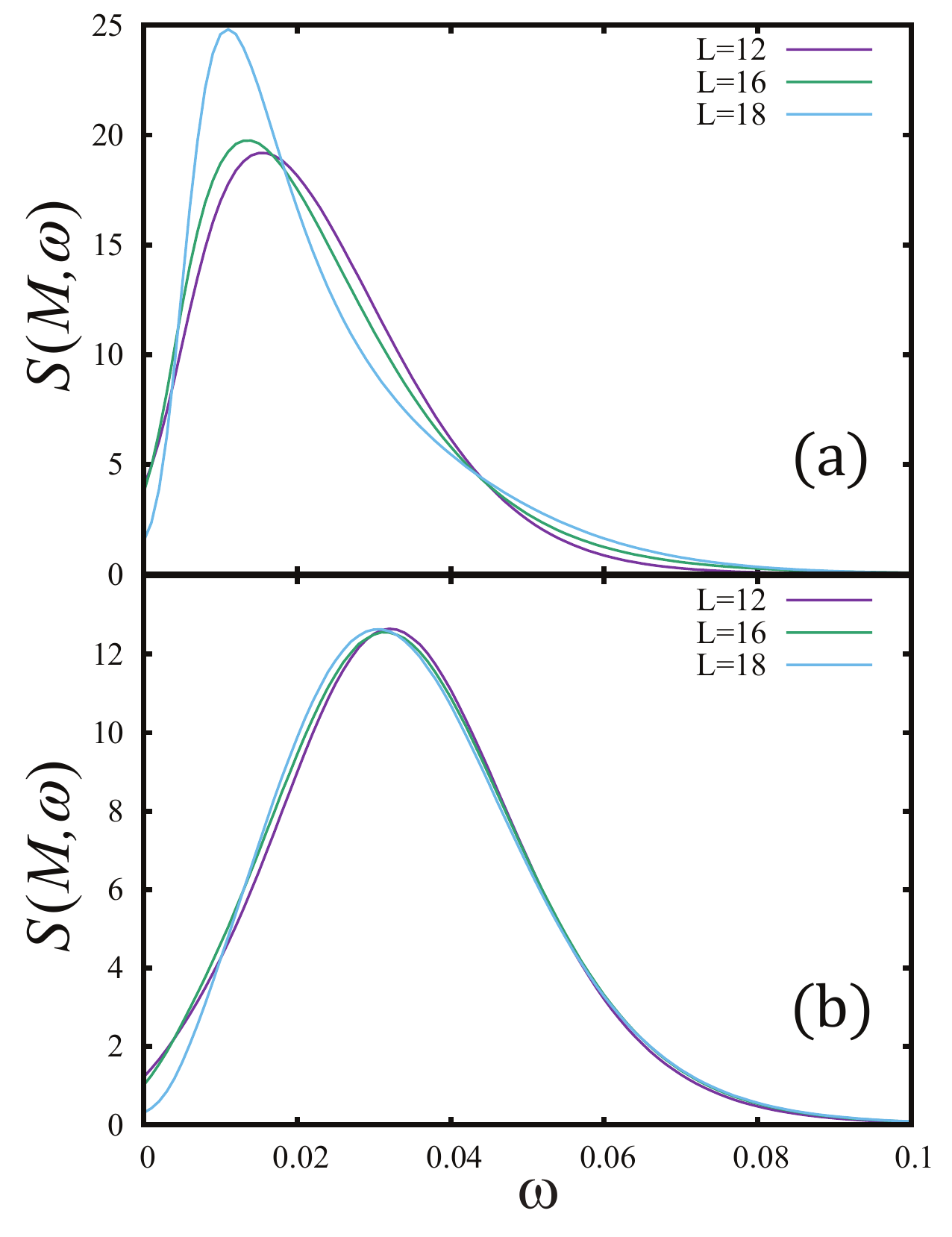}
\caption{(a) $S^{zz}(M,\omega)$ for 0-magnetization with $L=12,16$ and $18$. The parameters are the same those as shown in Fig.3 (a) of the main text. (b) $S^{zz}(M,\omega)$ for $1/6$-magnetization with $L=12,16$ and $18$. The parameters are the same as those as shown in the Fig.3 (b) of the main text.}
\label{fig:finitesize}
\end{figure}

Here we provide detailed information on the spectra close to QSL to VBS transition for both magnetizations. Fig.S1 (a) and (b) are the $S^{zz}(\omega)$ at $M$ point with different system size $L=12$, $16$ and $18$ for $m_z=0$ and $m_z=\frac{1}{6}$, respectively. The parameters are the same as those shown in the Fig.3 in the main text. Through the comparison of panels (a) and (b), one can see the position of spectral peak in (a) is apparently closer to $\omega=0$ than that in (b). Moreover, as the system size increases, at 0-magetization, the peak in (a) gradually moves towards $\omega=0$, which means the vison gap is closing in the infinite system. But for the spectra of $\frac{1}{6}$-magnetization in (b), the peak position of the peak does not change and the spectral weight is suppressing at $\omega=0$ with increasing $L$, which means there is a finite vison gap in the thermodynamic limit. These results are consistent with the dynamical signature of symmetry fractionalization for $0$-magnetization shown in the main text that the vison gap at $M$ point is closed while for $\frac{1}{6}$-magnetization the gap is still finite.

\section{Evolution of the dynamical spin structure factor}
In this session, we demonstrate the evolution of the DSSF at $m_z=\frac{1}{6}$ and $0$, across the FM to QSL phase transition.

\begin{figure}[htp!]
\centering
\includegraphics[width=\columnwidth]{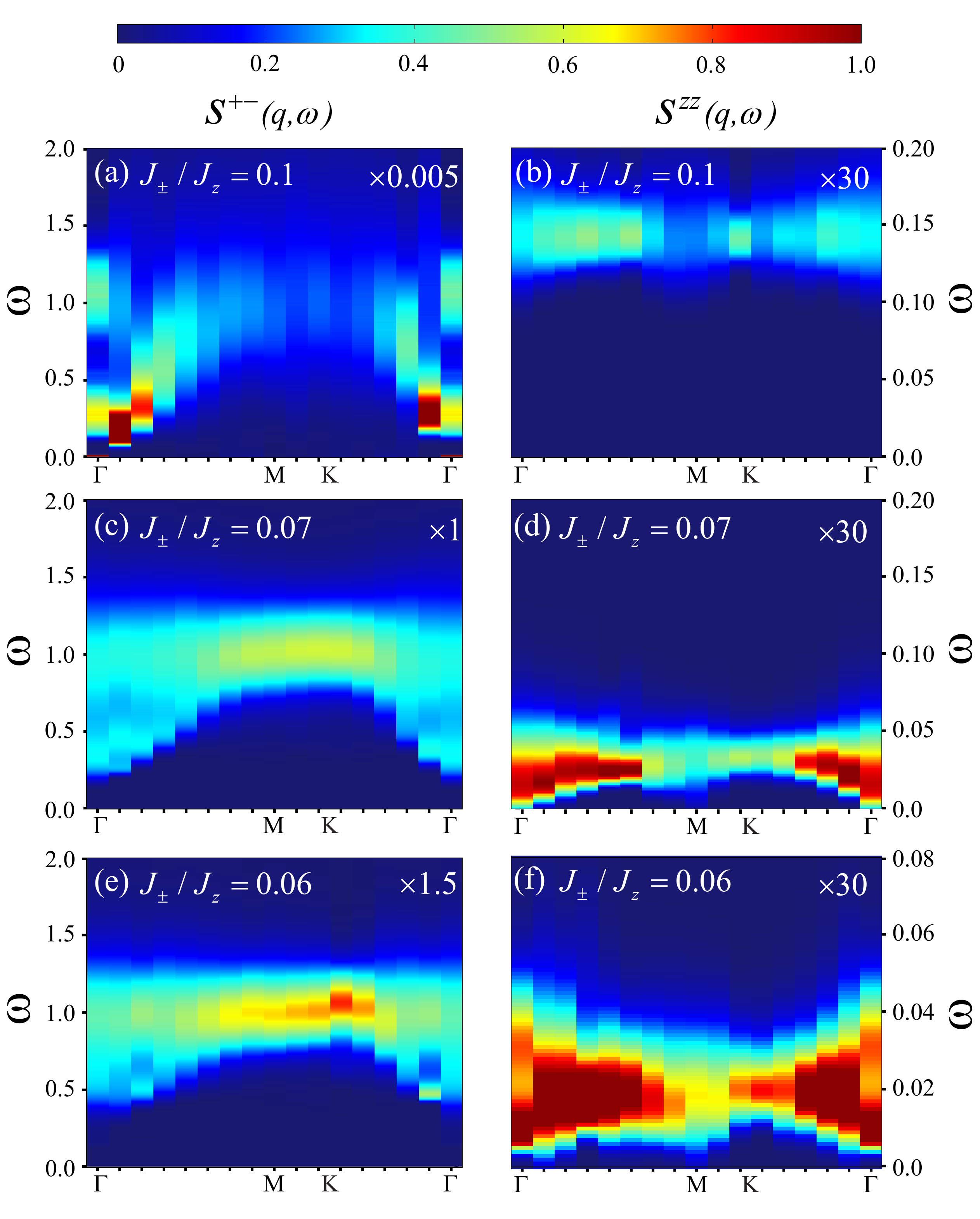}
\caption{DSSF for $m_z=\frac{1}{6}$, $L=16$ and $\beta=2L/J_{\pm}$ at $J'_z=0.005$ and varying $J_{\pm}/J_z$ across the FM to QSL phase transition.}
\label{fig:1/3filling}
\end{figure}

Fig.~\ref{fig:1/3filling} depicts the DSSF $S^{\pm}(\mathbf{q},\omega)$ and $S^{zz}(\mathbf{q},\omega)$ across the FM to QSL phase transition at $m_z=\frac{1}{6}$. The system size is $16\times 16$ and $\beta=2L/J_{\pm}$ as discussed in the main text. Fig. ~\ref{fig:1/3filling}(a), (b) are inside the FM phase and the Goldstone mode in the $S^{\pm}(\mathbf{q},\omega)$ close to $\mathbf{q}=\Gamma$ is clearly observed. $S^{zz}(\mathbf{q},\omega)$ is gapped for all the $\mathbf{q}$. Fig. ~\ref{fig:1/3filling}(c) and (d) are close to the phase transition and it is clear that the Goldstone mode in $S^{\pm}(\mathbf{q},\omega)$ is giving away to a gapped spectrum. In the $S^{zz}(\mathbf{q},\omega)$ channel, the gap becomes smaller close to $\mathbf{q}=\Gamma$ and the clear continuum of the vison pair excitations begins to form. Finally (e) and (f) are inside the QSL phase, the $S^{\pm}(\mathbf{q},\omega)$ is completely gapped, with the spinon pair continuum above the gap, and the vison continuum persists with a small vison gap.

\begin{figure}[htp!]
\centering
\includegraphics[width=\columnwidth]{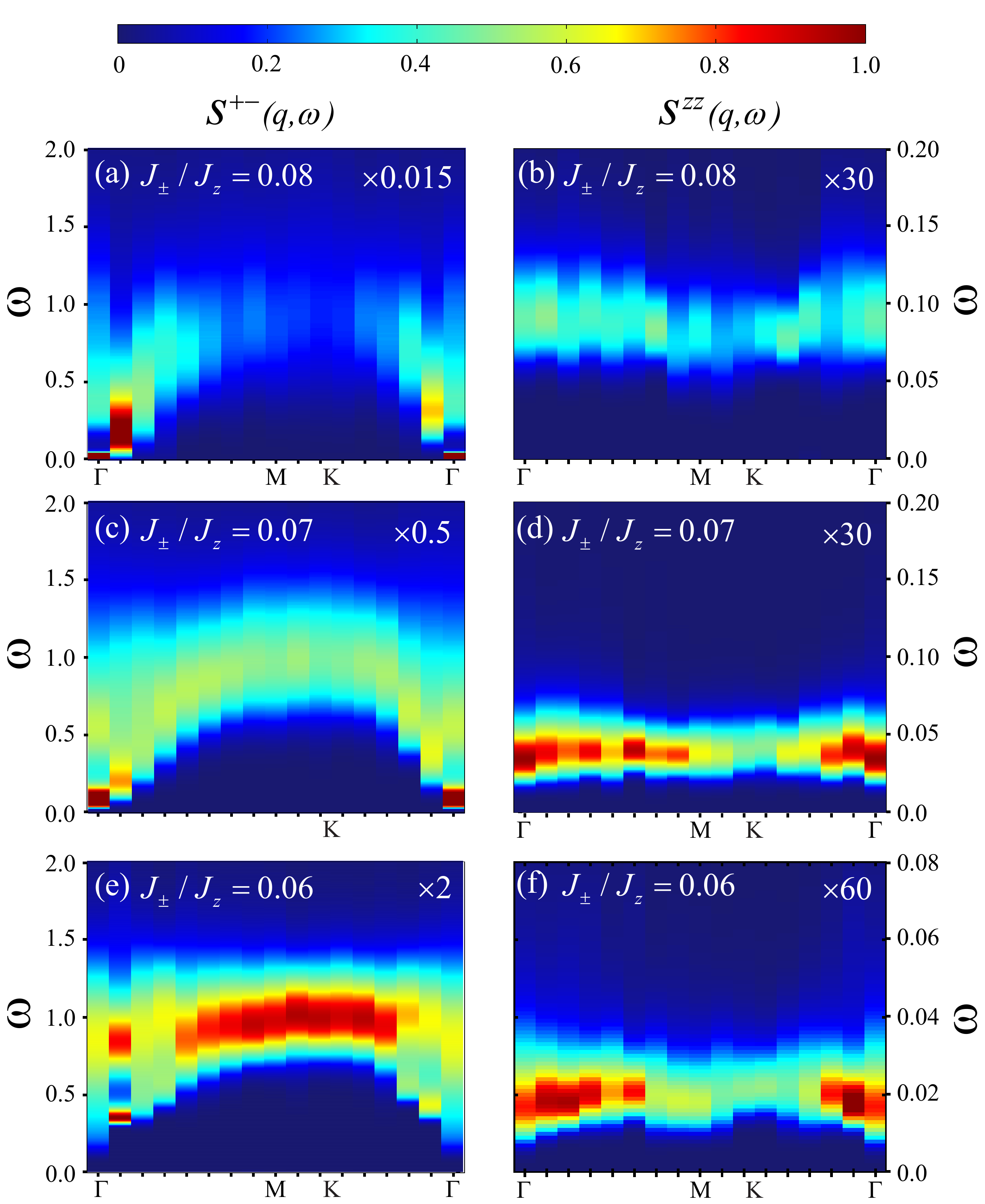}
\caption{DSSF for $m_z=0$, $L=16$ and $\beta=2L/J_{\pm}$ at $J'_z=0$ and varying $J_{\pm}/J_z$ across the FM to QSL phase transition.}
\label{fig:1/2filling}
\end{figure}

Fig.~\ref{fig:1/2filling} depicts the DSSF $S^{\pm}(\mathbf{q},\omega)$ and $S^{zz}(\mathbf{q},\omega)$ across the FM to QSL phase transition at $m_z=0$. The system size is $16\times 16$ and $\beta=2L/J_{\pm}$ as discussed in the main text. The phenomenology is completely analogous to the $m_z=\frac{1}{6}$ case discsussed earlier in Fig.~\ref{fig:1/3filling}, but with the difference that, as discussed in the main text, the translational symmetry fractionalization of the vison pair spectra are present in $S^{zz}(\mathbf{q},\omega)$ inside the QSL phase at $m_z=0$ only, so in Fig.~\ref{fig:1/2filling} (f), there is signature of the doubling of the period at the bottom of the continuum. And the doubling of the period is more pronounced as one drives the QSL toward VBS phase with $J'_z$, as shown in the Fig.3 of the main text.